\documentstyle[11pt,aaspp4]{article}
\newcommand\beq{\begin{equation}}
\newcommand\eeq{\end{equation}}

\def\unsetyr{\def\oyear{\relax}\def\cyear{\relax}\def\cyeara{a\relax}\def\cyearb{b\relax}\def\cyearc{c\relax}\def\cyeard{d\relax}}
\def\setyr{\def\oyear{(}\def\cyear{)}\def\cyeara{a)}\def\cyearb{b)}\def\cyearc{c)}\def\cyeard{d)}}
\unsetyr
\def\jcite#1{\setyr\cite{#1}\unsetyr}

\def\rmmat#1{{\hbox{\rm #1}}}
\def\rmscr#1{\rmmat{\scriptsize #1}}
\newcommand{\be}{\begin{equation}}
\newcommand{\ee}{\end{equation}}
\newcommand{\bt}{\begin{table} \begin{center}}
\newcommand{\et}{\end{center} \end{table}}
\newcommand{\ba}{\begin{eqnarray}}
\newcommand{\ea}{\end{eqnarray}}

\newcommand{\eg}{{\it e.g.~}}

\newcommand{\comment}[1]{\relax}
\def\eqref#1{Equation~\ref{eq:#1}}

\begin{document}

\title{X-ray emission from middle-aged pulsars}
\author{Rosalba Perna$^{1,3}$, Jeremy Heyl$^{2,3}$ \& Lars Hernquist$^3$}
\medskip
\altaffiltext{1}{Harvard Junior Fellow}
\altaffiltext{2}{{\em Chandra} Fellow}
\altaffiltext{3}{Harvard-Smithsonian Center for Astrophysics, 60 Garden Street,
Cambridge, MA 02138}

\begin{abstract}
We present a simple, unified model which accounts for properties of
the X-ray emission from the three middle-aged pulsars PSR 1055-52,
PSR 0656+14 and PSR 0630+18 (Geminga).  The X-ray radiation from
these objects is pulsed more strongly at energies above a transition point
around 0.5~keV.  In addition, the phase of the pulses shifts by about
$80^\circ-100^\circ$ degrees around the same point.  Geminga also has the
peculiarity that its pulsed fraction {\em decreases} in
the 0.3-0.5 keV energy range, attaining a minimum near 0.5~keV.  We
show that a two-component hydrogen atmosphere is able to account for
these disparate features.  In our model, the hotter component is powered 
by particle
bombardment and is restricted to the polar regions, while
the cooler one covers the entire stellar surface.  The
two components also differ in their emission patterns, with the
hard and soft contributions coming from areas radiating into
fan and pencil beams, respectively.
\end{abstract}


\section{Introduction}

X-ray emission from cooling neutron stars (NSs) had been predicted
and studied in detail already in the early 1960's (\cite{Chiu64}; 
\cite{Tsur64}).
However, it was not until the late 70's that the
{\em Einstein} Observatory first detected radiation from
isolated NSs, and until the {\em ROSAT} era (the 1990's) that more
detailed information on the spectral characteristics and pulsed
fraction ($Pf$) could be obtained.

Thermal radiation due to cooling can be best observed in a small
subset of  pulsars which are
middle-aged and have characteristic lifetimes $\tau\sim
10^5$ yr.  In younger pulsars, such as the Crab, X-rays of
magnetospheric origin often dominate over thermal emission, while
older ones are too cold to have any observable radiation due to
cooling. So far, the three best candidates for which
thermal emission has likely been detected
are PSR 1055-52, PSR 0656+14 and PSR 0630+18.

Apart from a non-thermal, power-law component dominating the
high-energy tail (this is particularly evident in PSR 0656+14; \cite{Grei96})
and which is thought to be
due to magnetospheric processes, the bulk of the emission
between $\sim 0.1-1$ keV is well fitted by a double blackbody at two
different temperatures.  The ratio of
the area of the hotter component to the
colder one is typically very small ($\sim$ a few $\times
10^{-5}$ to a few $\times 10^{-3}$).  The colder component has
been interpretated as being due to thermal cooling, while the hotter
one most likely arises from heated polar caps, with the extra heating being
due to bombardment by high-energy particles (\cite{Grei96}).

An important property of the radiation from all three pulsars (for a
review see Becker \& Trumper 1997) is a phase shift of $\sim 80^\circ -
100^\circ$ at a transition point which is around 0.5-0.6 keV. The
pulsed fraction also shows transitions in its behavior in this
region for both PSR 1055-52 and PSR 0656+14. Below $\sim 0.5$ keV, it is
roughly constant at the 10\% level, while it rapidly increases after
that point to about 20-30\% for PSR 0656+14 and to about 80-90\% for
PSR 1055-52. On the other hand, the pulsed fraction from Geminga shows
an intriguing feature: in the PSPC channels 8-28 (i.e. roughly at
energies below 0.3 keV) the $Pf$ is much larger than in channels 28-53
(roughly corresponding to the energy range 0.3-0.5 keV): 33\% versus
20\%.  However, as discussed by \jcite{Page95} and \jcite{Page96},
no matter what the surface temperature distribution is, as
long as it is not uniform, blackbody emission always gives an increase
of the $Pf$ with energy. Detailed modeling by \jcite{Shib95}
shows that realistic atmospheres are able to produce a
slight decrease of the $Pf$ with increasing energy, but it is still much
smaller than what is observed.  A model which could account for the
large observed decrease has been presented by \jcite{Page95b}. They
assume the presence of warm, magnetized
plates on the surface of Geminga, surrounded by cold, 
unmagnetized regions, with the warmer
plates emitting a softer spectrum than the surrounding areas. In
this model, Geminga would therefore have different characteristics
than the two other objects in the same class.

In this {\em Letter}, we present a simple model for the emission in
the $0.1\sim 1$ keV range that is able to account for the properties
of all three pulsars within a unified framework.  We model the cooling
component as a blackbody modified by a light-element atmosphere; as
discussed in the literature (\cite{Heyl98rxj,Zavl96,Pavl94}),
this radiation, which
emerges through an atmosphere heated from below, is most likely emitted in
a ``pencil'' beam (\cite{Pavl94,Zavl96}).  The hotter component, 
due to particle
bombardment of the polar caps, is produced in an atmosphere that is hotter
on top, i.e. at lower densities. This is due to the fact that
the tenuous upper reaches of the neutron-star atmospheres are
extremely inefficient emitters; therefore, as the bombarding particles
deposit energy in this region, it heats up dramatically to maintain
radiative balance.  Meanwhile, the denser layers of the atmosphere
which have a larger opacity heat up relatively little.
The problem of atmospheres heated from above has been discussed in
the literature in the context of slow accretion onto neutron stars
(\cite{2000ApJ...537..387Z}) and the X-ray illumination of normal stars in
X-ray binaries (\eg\ \cite{1975ApJ...196..583M}).
It has been shown that the radiation from an atmosphere illuminated
from above may be emitted in a ``fan'' pattern (\eg\ \cite{1975ApJ...196..583M}).
In fact, as one moves away from the normal, the effective photosphere of the
atmosphere moves to lower densities due to geometrical effects.   In a bombarded
atmosphere, these lower densities are hotter, so one sees a hotter and
brighter spectrum as one approaches grazing incidence.

We show that a combination of pencil beaming from the cooling
component and fan beaming from the hotter polar caps is able to
account not only for the magnitude of the pulsations in PSR 1055-52
and PSR 0656+14, but also for the phase shift between the soft and the
hard components. Moreover, this same model is able to account also for
the {\em decrease} of the pulsed fraction observed in the soft X-ray
emission of Geminga, without requiring any special model for the
composition of the surface of this object.

The paper is organized as follows: in \S2, we describe in detail the
X-ray spectral model; specific applications to PSR 1055-52, PSR
0656+14 and Geminga are discussed in \S 3; our results are summarized
and discussed in \S 4.

\section{X-ray spectrum}

We model the local emission from the surface of the star, $n(E,T)$,
as a blackbody spectrum modified
by the presence of an atmosphere, for which we adopt the semianalytical
model developed by \jcite{Heyl98atm} with a geometric generalization to
study the spectral intensity away from the normal.
The dependence of our results on the
atmospheric composition will be discussed in \S 3.   
We assume that the neutron star has
a dipolar magnetic field. With this field geometry and a sufficiently
strong intensity, the flux transmitted through the envelope can be
approximated as $F\propto\cos^2\psi$, where
$\cos^2\psi=4\cos^2\theta_p/(3\cos^2\theta_p+1)$ is the angle between
the radial direction and the magnetic field (\cite{Gree83};
\cite{Heyl97analns}, 2000); here
\beq
\theta_p=\arccos(\cos\theta\cos\alpha+\sin\theta\sin\alpha\cos\phi)
\label{eq:tetap}
\eeq
is the angle that the magnetic pole makes with the normal to the
star at position $(\theta,\phi)$, while $\alpha$ is the angle that it
makes with the line of sight. If $\xi$ is the angle between the 
magnetic dipole and the rotation axis, and $\chi$ the angle between
the observer's direction and the rotation axis, then the angle $\alpha$
is given by 
\beq
\alpha=\arccos(\cos\chi\cos\xi+\sin\chi\sin\xi\cos\gamma)\;,
\label{eq:alpha}
\eeq
with $\gamma$ being the phase angle.

Outside of the heated polar caps, the local temperature on the star
due to thermal cooling is given by 
\beq
T_{\rm th}(\theta,\phi)=T_p\left[\frac{4\cos^2\theta_p}{3\cos^2\theta_p+1}\;
(0.75\;\cos^2\theta_p+0.25)^{0.2}\right]\;,
\label{eq:Tdip}
\eeq 
where, following \jcite{Heyl97analns}, we have assumed a
further dependence of the flux on magnetic field strength
scaling as $B^{0.4}$. In Equation (3), 
$T_p$ is the temperature
that the pole would have if there were no reheating. 

Now, let $\beta$ be the angular size of the radius of the polar caps, which are centered
around the poles. The hot spot region (in the upper hemisphere)
is defined by 
\beq
\theta\le\beta,\;\;\;\;\;\;\;\;\;\;\;\; \rmmat{if}\;\;\; \alpha=0\;
\label{eq:con1}
\eeq
and
\beq
   \left\{
  \begin{array}{ll}
    \alpha-\beta\le\theta\le\alpha+\beta \\
      \phi\le\phi_p \;\;\;\rmmat{or}\;\;\;\; 2\pi-\phi\le\phi_p, \;\;\;\;\rmmat{if}
        \;\;\;\alpha\ne 0\;\;\;\rmmat{and} \;\;\;\beta\le\alpha\\
  \end{array}\right.\;
\label{eq:con2}
\eeq
where
\beq
\phi_p=\arccos\left[\frac{\cos\beta-\cos\alpha\cos\theta}{\sin\alpha\sin\theta}\right]\;.
\label{eq:phip}
\eeq
Here, we have chosen a system of coordinates such that the prime
meridian passes through the North pole of the star, at the center of
the upper polar cap.  The hot spot in the lower hemisphere is defined
by the condition $\theta\le\pi-\beta$, if $\alpha= 0$, and by
Equation (5) and Equation (6) with the
substitutions $\alpha \rightarrow \pi-\alpha$ and
$\phi\rightarrow\pi-\phi$, if $\alpha\neq 0$. 

Inside the hot spot, we assume the temperature to be constant, $T_\rmscr{hs}(\theta,\phi)
={\rmmat const} \equiv T_{\rm hs}$. As already discussed, 
we adopt pencil beaming for the thermal component and fan beaming for the
radiation produced in the polar caps, and, for simplicity,  we parameterize them with
the functions
\beq
   \left\{
  \begin{array}{ll}
f_\rmscr{th}(\delta)\propto \cos^{n_1}(\delta)\\
f_\rmscr{hs}(\delta)\propto \sin^{n_2}(\delta)\\
  \end{array}\right.\;,
\label{eq:beam}
\eeq
where $\delta$ is the angle that a photon emitted at a colatitude
$\theta$ on the star makes with the normal to the surface at the
moment of emission. The relation between $\theta$ and $\delta$ 
(which is a consequence of the general relativistic effects
of light deflection) is
given by the ray-tracing function (Page 1995)
\beq
\theta(\delta)=\int_0^{R_s/2R}x\;du\left/\sqrt{\left(1-\frac{R_s}{R}\right)
\left(\frac{R_s}{2R}\right)^2-(1-2u)u^2 x^2}\right.\;,
\label{eq:teta}
\eeq having defined $x\equiv\sin\delta$.  The value of $n_1$ in the
above equation is found to be roughly unity for an unmagnetized
light-element atmosphere (\cite{Zavl96}) and slightly larger for a
magnetized light-element atmosphere (\cite{Pavl94}), while it is about
0.5 for an atmosphere made of heavy elements (\cite{Raja97}).  In our
calculations we restrict ourselves to this range of values.  The
exponent $n_2$, on the other hand, is treated as a fit parameter and
is allowed to vary from source to source, as one would expect the
extent of fan beaming in the bombarded atmosphere to depend on the
spatial and energy distribution of the bombarding particles which are
unknown (and likely to vary among different objects).

The calculation of the emitted spectrum fully accounts for the
consequences of gravitational bending of light
and gravitational redshift.  Let us define
$e^{-\Lambda_s}=\sqrt{1-R/R_s}$, where $R$ is the radius of the neutron
star, and $R_s=2GM/c^2$ is its Schwarzschild radius.  A range of radii
compatible with the currently available models for the NS equation of
state requires $2\le (R/R_s)\le 4$. Here we take $R$ in this range, with $M=1.4
M_\odot$.  Let $D$ be the distance from the star to the observer, and
$N_\rmscr{H}$ the intervening column density.  The flux measured by an
observer at infinity is then given by the sum of the contributions
from the thermal component and from the polar caps
\begin{eqnarray}
F(E;\gamma)&=&\frac{\pi R^2\;e^{-\Lambda_s}}
{4\pi D^2}e^{-\sigma(E)N_\rmscr{H}}
\int_0^1 2xdx\int_0^{2\pi}\frac{d\phi}{2\pi}\nonumber \\ &\times& \left\{ 
\frac{1}{k T_{p}}\sigma T^4_\rmscr{th}(\theta,\phi) \;
n[Ee^{-\Lambda_s};T_\rmscr{th}(\theta,\phi)]\;+\;
\frac{1}{k T_\rmscr{hs}}\sigma T^4_\rmscr{hs}(\theta,\phi) \;
n[Ee^{-\Lambda_s};T_\rmscr{hs}(\theta,\phi)]\right\}\;,
\label{eq:flux}
\end{eqnarray}
in units of phot cm$^{-2}$ s$^{-1}$ keV$^{-1}$. Note that the dependence 
on $\gamma$ comes in through Eqs. (1) -- (3).

Finally, the energy dependent pulsed fraction is defined by
\beq
Pf(E)=\frac{F^\rmscr{max}(E)-F^\rmscr{min}(E)}
{F^\rmscr{max}(E)+F^\rmscr{min}(E)}\;,
\label{eq:pf}
\eeq
where $F^\rmscr{max}(E)$ and $F^\rmscr{min}(E)$ are, respectively, the maximum
and minimum flux over a rotation period of the star.  The phase of the
pulsation, $Ph(E)$, is defined as the phase angle $\gamma$ at which the flux 
is maximum at energy $E$. 

\section{Application to PSR 1055-52, PSR 0656+14 and PSR 0630+18}

For each pulsar, we considered three types of models: a two-component
blackbody (of the type expected from iron atmospheres),
or a two-component light-element atmosphere, and for the
latter we considered two different types of opacities.  The emergent
spectrum from iron atmospheres is similar to a blackbody in its gross
properties (\cite{Raja97}) and we assume that it is moderately beamed
as discussed in \S 2.  The two-component light-element atmosphere is
substantially bluer than a blackbody and also exhibits a hard tail.
We have used two different semianalytic atmospheres of
\jcite{Heyl98atm} with opacities decreasing as $\nu^{-3}$ and
$\nu^{-1}$.  The former model is much bluer than a blackbody and
exhibits a strong hard tail.  
The latter is most appropriate to model magnetized light-element
atmospheres (\eg\ \cite{Pavl94}); it is intermediate between the
$\nu^{-3}$-model and the blackbody.

The effective temperatures (parameterized through $T_p$ and $T_{\rm
hs}$) and the size of the hot region (called $\beta$ in our paper)
correponding to the same effective blackbody temperatures are
different in the three models, and, to precisely determine them in
each case, we generated a simulated spectrum (using the software
XSPEC, \cite{A96}, and the $ROSAT$ PSPC detector) of each object,
using the temperature values of the double-blackbody fits and the
ratio between the areas of the two components given in the literature.
This spectrum was then fitted with the phase averaged spectrum of our
model\footnote{This is given by $\frac{1}{2\pi} \int_0^{2\pi}d\gamma
F(E;\gamma)$.}  to determine $T_p$, $T_{\rm hs}$ and $\beta$. These
values were then used to compute pulsed fractions and phases with some
choices of $n_1$ and $n_2$.

We found that an iron-type atmosphere, which has a rather mild degree
of beaming, was unable to reproduce the degree of modulation observed
in the softest part of the spectrum (dominated by the thermal
component\footnote{Note, however, that this result is dependent on the
assumption of a dipolar field. The value of the pulsed fractions can be
increased if a quadrupole component is added (\cite{Page96}).}). 
A light element atmosphere with an opacity $\propto\nu^{-3}$ was able
to reproduce reasonably well the pulsed fractions in the soft component
of the spectrum. However, the tail of the thermal component
is generally so strong that a
second hotter component was not statistically required by the data
in the case of PSR 0656 and PSR 1055, and it was much smaller
than the one required by the double blackbody fit in the case of Geminga\footnote{
Note that in the case of Geminga, similar results had been found by \jcite{MPM94}.}.
Such a model was unable to produce the sudden transition in the
behaviour of the pulsed fractions accompanied by a phase shift around
0.5 keV which is observed in PSR 0656 and PSR 1055. 
Moreover, such models implied surface areas much larger than what would
be consistent with any reasonable equation of state for the neutron star. 
A two-component atmosphere model which is less blue ($\propto\nu^{-1}$) 
than the former required a second hotter component in the spectral fitting,
and was able to reproduce the behaviour of the observed pulsed fractions and
phase shifts. More details on each of the object that we considered are
given in the following. 

\subsection{PSR 1055-52}

This pulsar has a characteristic age $\tau=P/2\dot{P}\sim 5\times
10^5$ yr, and a rotational energy loss rate of $\dot{E}\sim 3\times
10^{34}$ erg/sec. {\em ROSAT} observations (\cite{Ogel93a}) clearly
show that the pulses have an energy dependent phase and pulsed
fraction, with a transition at $\sim 0.5$ keV. The pulsed fraction
below the transition is roughly constant at a value on the order of
7-8\%, while after the transition it rapidly increases to a maximum of
0.85 around 1 keV. The hard photons lead the soft ones by an angle
$\sim 120^\circ$.  A blackbody fit to the spectrum requires two
components, with temperatures $T_\rmscr{soft} \sim 8 \times 10^5$ K
and $T_\rmscr{hard}\sim3.7\times 10^5$ K, and a ratio between the 
areas of $A_\rmscr{soft}/A_\rmscr{hard}\sim 3\times 10^{-5}$
(\cite{Grei96}).  The angles $\xi$ and $\chi$ have been inferred by
\jcite{Malo90} to be both about $30^\circ$. With these values, and
$T_\rmscr{p}$, $T_\rmscr{hs}$ and $\beta$ calibrated on the phase averaged spectrum,
we show the derived pulsed fractions and phase for this pulsar in Figure
1, having assumed $n_1=1$ and $n_2=2$.

\subsection{PSR 0656+14}

PSR 0656+14 was discovered in an {\em Einstein} satellite survey of
ultrasoft sources (\cite{1989ApJ...345..451C}). It has a relatively
young spin down age of $1.1\times 10^5$ yr and a rotational energy
loss rate $\dot{E}\sim 3.8 \times 10^{34}$ erg/sec. The combined {\em
ASCA} and {\em ROSAT} spectrum reveals two blackbody components with
$T_\rmscr{soft}\sim 8\times 10^5$ K, and $T_\rmscr{hard}\sim 1.5\times
10^6$ K, and shows evidence that a power-law component is needed to
account for higher energy photons. The ratio of the hot polar cap
area to the neutron star surface area is $\sim 7\times 10^{-3}$
(\cite{Grei96}).

As for PSR 1055-52, both the phase and the pulsed fraction are energy
dependent. Below $\sim 0.5$ keV, the pulsed fraction stays a little under
$\sim 10\%$, and it increases modestly in the range $0.5-1$~keV
to around
20-30 \% at 1 keV. The soft and the
hard components are shifted in phase by about $85^\circ$ with respect
to each other.  The angles $\xi$ and $\chi$ have been estimated by
\jcite{Malo90} to be both about $35^\circ$. With these values, and
$T_\rmscr{p}$, $T_\rmscr{hs}$ and $\beta$ derived from spectral fitting,
we show the pulsed fraction and phase that
our model predicts with the choice $n_1=0.8$ and $n_2=0.2$ in Figure 2.

\subsection{PSR 0630+18 (Geminga)}

Geminga was first observed as a high-energy $\gamma$-ray source with
the SAS-2 satellite in the 100 MeV band 
(\cite{Fich75}). It was
only in 1992 that it was observed in the X-ray band with {\em
ROSAT} (\cite{Halp92}).  Its period and spin-down rate (\cite{Bert92}; 
\cite{Herm92})
 yield a dynamic age of $3.4\times
10^5$ yr, and a rotational energy loss rate $\dot{E} \sim 3.5\times
10^{35}$ erg/sec. A spectral analysis of the {\em ROSAT} data made
by \jcite{Halp93} shows that the X-ray spectrum consists
of two blackbody components with $T_\rmscr{soft}\sim 5\times 10^5$ K,
and $T_\rmscr{hard}\sim 3\times 10^6$ K. Both components are modulated
at the pulsar rotation period, but the harder X-ray pulse leads 
the soft pulse by about $105^\circ$ in phase. Geminga is believed to
be an orthogonal rotator, with $\xi=\chi\sim 90^\circ$ (\cite{Malo90}). 

Unlike PSR 1055-52 and PSR 0656+14, the pulsed fraction observed in
Geminga shows a decrease with energy at low energies: the amplitude of
the pulsations in the PSPC channels 8-28 (i.e. roughly at energies
below 0.3 keV) is much larger than in channels 28-53 (roughly
corresponding to the energy range 0.3-0.5 keV): 33\% versus 20\%.
Figure 3 shows that a significant decrease of the pulsed
fractions in the 0.3-0.5 keV energy range can be reproduced with the
model described in \S 2: here we have taken $n_1=1.5$ and $n_2=3$,
while all other model parameters have again been calibrated from a fit
to the phase-averaged spectrum\footnote{Note that a similar behaviour
for the $Pfs$ in the softest part of the spectrum 
could be obtained with a more modest beaming (i.e. smaller $n_1$), but
a quadrupole component of the $B$ field added to the dipole (\cite{}Page96).}.  
The decrease in the pulsed fraction
occurs in the region where the harder component starts to overtake the
softer one. The hard component brings more photons at phase angles in
which the soft component has less photons, therefore leading to a
decrease of the intensity fluctuations produced by the thermal
component alone.  This effect happens in the same way also in the
other two examples that we showed, but due to the less intense beaming
of the harder component assumed there, the decrease is not so
pronounced.  Interestingly, a slight decrease in the pulsed fraction
before its rapid increase can be observed also in the data for these
other two pulsars (e.g. \cite{Ogel95}).

Our model for the phase shifts and pulsed fraction in Geminga is
similar to that proposed by \jcite{Halp93}.  \jcite{Page95b} later
argued that this picture could not work because the flux in the hard
component is more than an order of magnitude less than that in the
soft component in the $0.3-0.5$ keV energy range, where the decrease  
in the $Pf$ is observed. However, the
pencil-beaming reduces the flux of the soft component at the same
phase of the rotation where the fan-beaming of the hard component
increases its flux; therefore, although the phase averaged fluxes in
the two components are not comparable, the instantaneous fluxes 
at $\gamma \sim 90^\circ$ can be, 
and the hard component can begin to influence the soft one well before
it becomes dominant in the phase-averaged spectrum.  

\section{Summary and Discussion}

We have proposed a simple model that is able to account for properties
of the X-ray emission observed in the three middle aged pulsars PSR
1055-52, PSR 0656+14 and Geminga within a unified framework.  The
emission in the $\sim 0.1-1$ keV energy range is believed to be due to
the combination of thermal radiation from the entire star and emission
from heated polar caps. We have argued that, while pencil beaming is
expected for the thermal, soft component, the hard component from the
polar caps is expected to be beamed into a fan, as discussed in the
literature for similar problems in other contexts.  We have shown that
such a model is able to account not only for the magnitude of the
pulsations in PSR 1055-52 and PSR 0656+14, but also for the phase
shifts between the soft and the hard component observed in all the
three objects.  We have shown that this same model is able to
reproduce also the decrease of the pulsed fraction observed in the
soft X-ray emission of Geminga, without requiring any special model
for the composition of the surface of this object. However, we have
found that the type and composition of the atmosphere plays an
important role. An atmosphere made of heavy elements is not able to
account for significant $Pfs$ (if only a dipolar field is assumed for
the thermal component), while a light element atmosphere with opacity
$\propto\nu^{-3}$ has a very pronounced tail which is able to account
for the all spectrum (up to about 1.5 - 2 keV) for PSR 0656 and PSR
1055, without requiring a second, hotter component.  Such a model is
not able to account for the sudden increase of the $Pfs$ (accompanied
by a phase shift) that is observed for these two pulsars around 0.5
keV. In the case of Geminga, the implied size of the hot spots is too
small to produce any significant effect.  On the other hand, a light
element atmosphere with opacity $\propto\nu^{-1}$ requires a second,
hotter component to account for the overall spectrum, and, as we have
shown, such a model is able to account for the observed pulsed
fractions and phase shifts, if one allows the emission from the hot
spots to have a variable degree of beaming for the various objects.
We have argued that this this is plausible, as it depends on the
spatial and energy distribution of the bombarding particles, which is
likely to vary among the various objects.

In conclusion, we need to stress that the model we have presented here was
aimed at explaining features observed in the X-ray emission from
middle aged pulsars from a qualitative point of view.  A more
quantitative analysis, which includes detector and absorption effects
\footnote{For a perfect (diagonal) detector, absorption
does not affect the {\em energy-dependent} $Pfs$ (but it does
affect the $Pf$ over a finite energy bandwidth; \cite{PHH2000}); however, for a real
detector, absorption can affect also the energy-dependent $Pfs$ to
a certain extent (\cite{Page95}).}  
would not change any of the qualitative features reproduced here. On
the other hand, a thorough analysis, inclusive of
detailed predictions for the shape of the pulse profiles, would require
very detailed models for the atmosphere and for the temperature
distribution on the star, as well as much better data with which to 
compare the model.  We anticipate that forthcoming
observational data
from the {\em Chandra} and {\em XMM} missions
will soon make this endeavor feasible.

\acknowledgements{We thank Jonathan McDowell for support with use of
 the XSPEC software}.

%\bibliographystyle{jer}
%\bibliography{ns,physics,mine,gr}

\begin{thebibliography}{}

\bibitem[\protect{Arnaud~\protect\oyear 1996\protect\cyear}]{A96}
\newblock
Arnaud, K. A. 1996, in ASP Conf. Series 101, {\em Astronomical
Data Analysis Software and Systems V}, ed. G. Jacoby \& J. Barnes
(San Francisco: ASP), 17

\bibitem[\protect{Bertsch et~al.~\protect\oyear 1992\protect\cyear}]{Bert92}
Bertsch, D.~L. {\it et~al.} 1992,
\newblock {\em Nature,} {\bf 357}, 306.

\bibitem[\protect{Becker \& Trumper~\protect\oyear 1997\protect\cyear}]{BT97}
Becker, W. \& Trumper, J. 1997,
\newblock {\em A\&A,} {\bf 326}, 682

\bibitem[\protect{Chiu \& Salpeter~\protect\oyear 1964\protect\cyear}]{Chiu64}
Chiu, H.-Y. \& Salpeter, E.~E. 1964,
\newblock {\em Phys. Rev. Lett.,} {\bf 12}, 413.

\bibitem[\protect{{Cordova} et~al.~\protect\oyear
  1989\protect\cyear}]{1989ApJ...345..451C}
{Cordova}, F.~A., {Middleditch}, J., {Hjellming}, R.~M. \& {Mason}, K.~O. 1989,
\newblock {\em \apj,} {\bf 345}, 451.

\bibitem[\protect{Fichtel et~al.~\protect\oyear 1975\protect\cyear}]{Fich75}
Fichtel, C.~E. {\it et~al.} 1975,
\newblock {\em ApJ,} {\bf 198}, 163.

\bibitem[\protect{Greenstein \& Hartke~\protect\oyear
  1983\protect\cyear}]{Gree83}
Greenstein, G. \& Hartke, G.~J. 1983,
\newblock {\em ApJ,} {\bf 271}, 283.

\bibitem[\protect{Greiveldinger et~al.~\protect\oyear
  1996\protect\cyear}]{Grei96}
Greiveldinger, C. {\it et~al.} 1996,
\newblock {\em ApJL,} {\bf 465}, 35.

\bibitem[\protect{Halpern \& Holt~\protect\oyear 1992\protect\cyear}]{Halp92}
Halpern, J.~P. \& Holt, S.~S. 1992,
\newblock {\em Nature,} {\bf 357}, 222.

\bibitem[\protect{Halpern \& Ruderman~\protect\oyear
  1993\protect\cyear}]{Halp93}
Halpern, J.~P. \& Ruderman, M. 1993,
\newblock {\em ApJ,} {\bf 415}, 286.

\bibitem[\protect{Hermsen et~al.~\protect\oyear 1992\protect\cyear}]{Herm92}
Hermsen, W. {\it et~al.} 1992,
\newblock IAU Circular No. 5541

\bibitem[\protect{Heyl \& Hernquist~\protect\oyear
  1998\protect\cyeara}]{Heyl97analns}
Heyl, J.~S. \& Hernquist, L. 1998a,
\newblock {\em MNRAS,} {\bf 300}, 599.

\bibitem[\protect{Heyl \& Hernquist~\protect\oyear
  1998\protect\cyearb}]{Heyl98rxj}
Heyl, J.~S. \& Hernquist, L. 1998b,
\newblock {\em MNRAS,} {\bf 297}, L69.

\bibitem[\protect{Heyl \& Hernquist~\protect\oyear
  1998\protect\cyearc}]{Heyl98atm}
Heyl, J.~S. \& Hernquist, L. 1998c,
\newblock {\em MNRAS,} {\bf 298}, L17.

\bibitem[\protect{Heyl \& Hernquist~\protect\oyear
  2000\protect\cyear}]{Heyl2000}
Heyl, J.~S. \& Hernquist, L. 2000,
\newblock {\em MNRAS}, in press 

\bibitem[\protect{Malov~\protect\oyear 1990\protect\cyear}]{Malo90}
Malov, I.~F. 1990,
\newblock {\em Astron. Zh.,} {\bf 67}, 377.,
\newblock Sov. Astron, 34, 189

\bibitem[\protect{Meyer et al.~\protect\oyear 1990\protect\cyear}]{MPM94}
Meyer, R. D., Pavlov, G. G., \& Meszaros, P. 1994,
\newblock {\em \apj} {\bf 433}, 265

\bibitem[\protect{{Milgrom} \& {Salpeter}~\protect\oyear
  1975\protect\cyear}]{1975ApJ...196..583M}
{Milgrom}, M. \& {Salpeter}, E.~E. 1975,
\newblock {\em \apj,} {\bf 196}, 583.

\bibitem[\protect{{\"{O}}gelman~\protect\oyear 1995\protect\cyear}]{Ogel95}
{\"{O}}gelman, H. 1995,
\newblock in M.~A. Alpar, U. Kiziloglu \& J.~V. Paradijs (eds.), {\em The Lives
  of Neutron Stars}, p. 101, Kluwer, Dordrecht

\bibitem[\protect{{\"{O}}gelman \& Finley~\protect\oyear
  1993\protect\cyear}]{Ogel93a}
{\"{O}}gelman, H. \& Finley, J.~P. 1993,
\newblock {\em ApJL,} {\bf 413}, 31.

\bibitem[\protect{Page~\protect\oyear 1995\protect\cyear}]{Page95}
Page, D. 1995,
\newblock {\em ApJ,} {\bf 442}, 273.

\bibitem[\protect{Page \& Sarmiento~\protect\oyear 1996\protect\cyear}]{Page96}
Page, D. \& Sarmiento, A. 1996,
\newblock {\em ApJ,} {\bf 473}, 1067.

\bibitem[\protect{Page, Shibanov \& Zavlin~\protect\oyear
  1995\protect\cyear}]{Page95b}
Page, D., Shibanov, Y.~A. \& Zavlin, V.~E. 1995,
\newblock {\em ApJL,} {\bf 451}, 21.

\bibitem[\protect{Pavlov et~al.~\protect\oyear 1994\protect\cyear}]{Pavl94}
Pavlov, G.~G., Shibanov, Y.~A., Ventura, J. \& Zavlin, V.~E. 1994,
\newblock {\em A\&A,} {\bf 289}, 837.

\bibitem[\protect{Perna et~al.~\protect\oyear 2000\protect\cyear}]{PHH2000}
Perna, R., Heyl, J., \& Hernquist, L. 2000, 
\newblock {\em ApJL,} {\bf 538}, 159.

\bibitem[\protect{Rajagopal, Romani \& Miller~\protect\oyear 1997\protect\cyear}]{Raja97}
Rajagopal, M., Romani, R.~W., \& Miller, M.~C. 1997, 
\newblock {\em ApJ,} {\bf 479}, 347.

\bibitem[\protect{Shibanov et~al.~\protect\oyear 1995\protect\cyear}]{Shib95}
Shibanov, Y.~A., Pavlov, G.~G., Zavlin, V.~E. \& Tsuruta, S. 1995,
\newblock in H. B{\"{o}}hringer, G.~E. Morfill \& J.~E. Tr{\"{u}}mper (eds.),
  {\em Seventeenth Texas Symposium on Relativistic Astrophysics and Cosmology},
  Vol. 759 of {\em Annals of the New York Academy of Sciences}, p. 291, The New
  York Academy of Sciences, New York

\bibitem[\protect{Tsuruta~\protect\oyear 1964\protect\cyear}]{Tsur64}
Tsuruta, S. 1964,
\newblock {\em Ph.D. thesis}, Columbia University

\bibitem[\protect{{Zane}, {Turolla} \& {Treves}~\protect\oyear
  2000\protect\cyear}]{2000ApJ...537..387Z}
{Zane}, S., {Turolla}, R. \& {Treves}, A. 2000,
\newblock {\em \apj,} {\bf 537}, 387.

\bibitem[\protect{Zavlin, Pavlov \& Shibanov~\protect\oyear
  1996\protect\cyear}]{Zavl96}
Zavlin, V.~E., Pavlov, G.~G. \& Shibanov, Y.~A. 1996,
\newblock {\em A\& A,} {\bf 315}, 141.

\end{thebibliography}

\newpage

\begin{figure}[t]
\centerline{\epsfysize=5.7in\epsffile{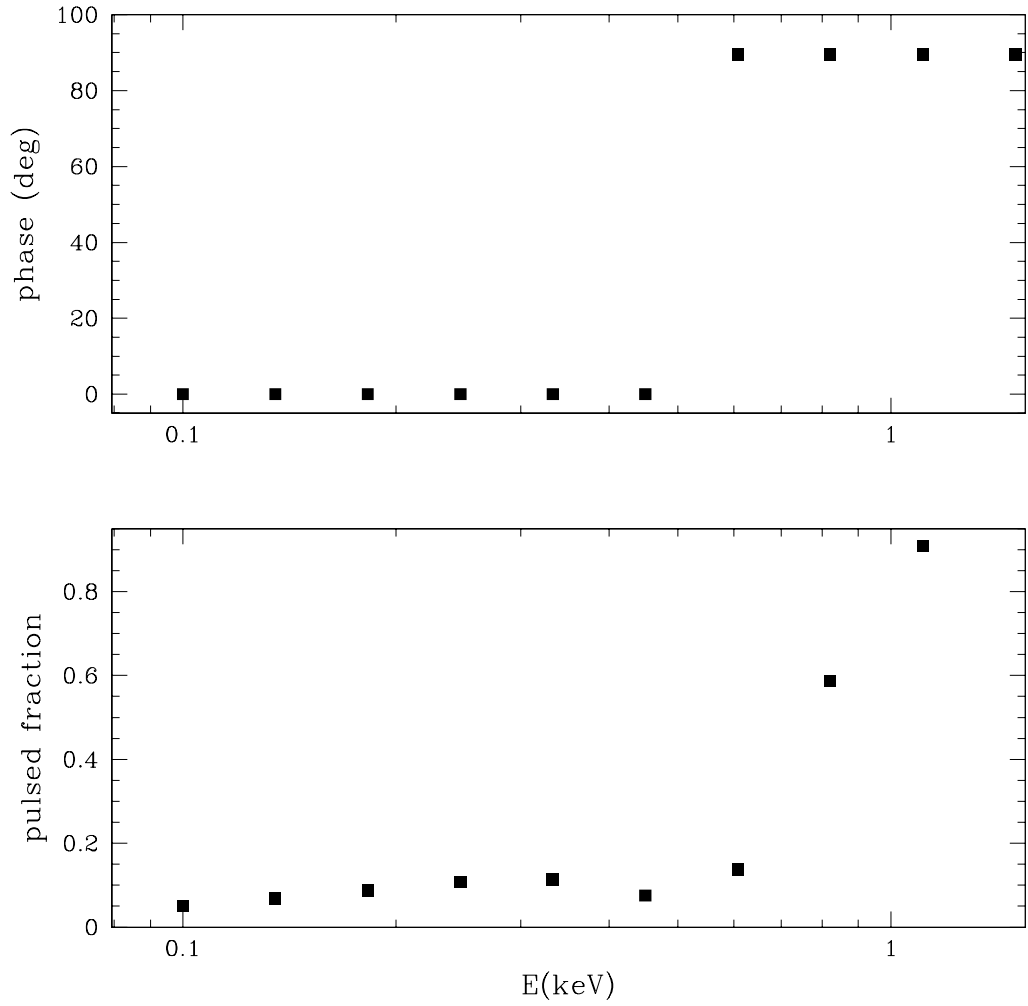}}
\caption{Predicted energy-dependent phase and pulsed fraction for
the X-ray emission from PSR 1055-52.}
\label{fig:1}
\end{figure}

\begin{figure}[t]
\centerline{\epsfysize=5.7in\epsffile{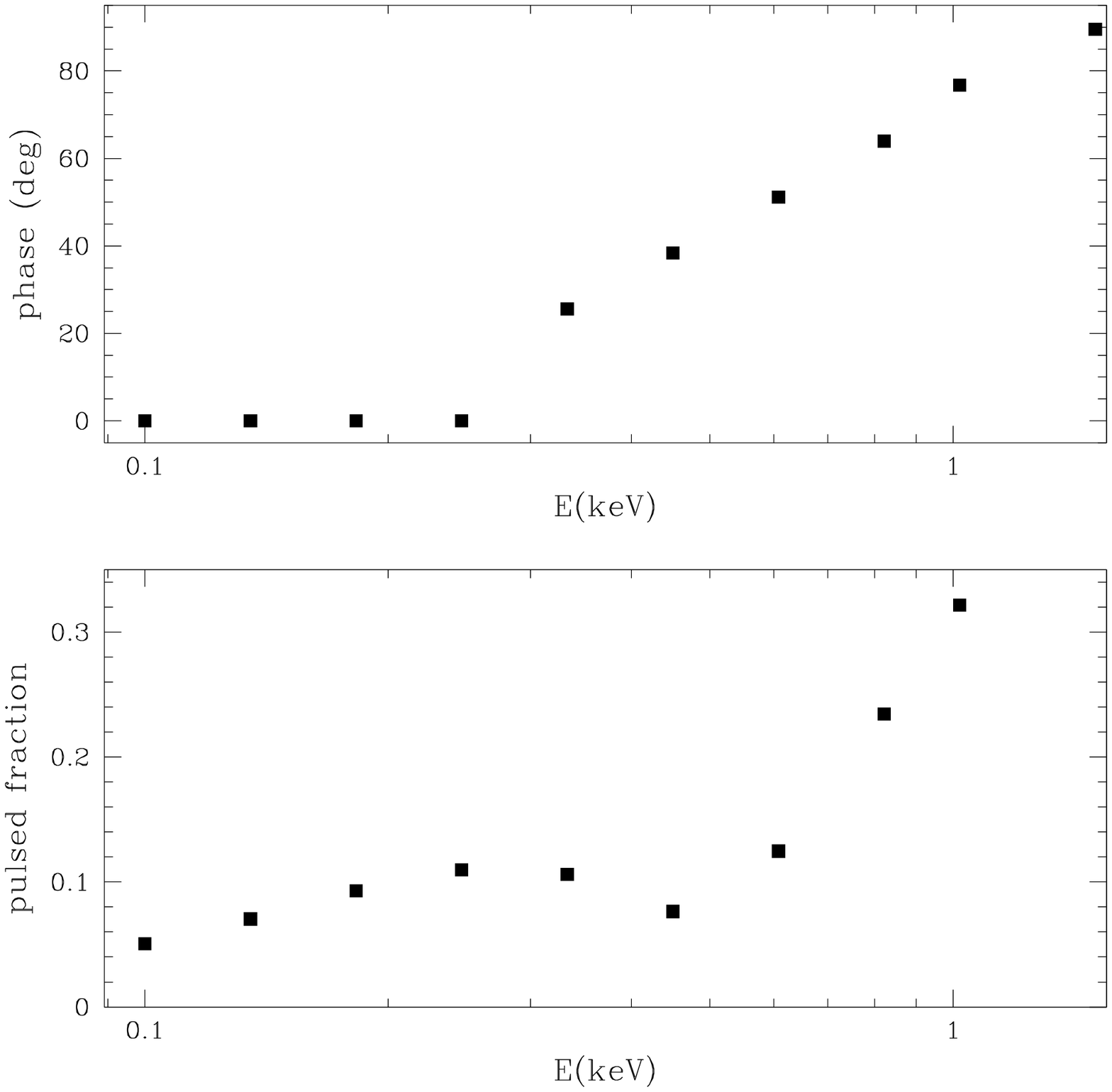}}
\caption{Predicted energy-dependent phase and pulsed fraction for
the X-ray emission from PSR 0656+14.}
\label{fig:2}
\end{figure}

\begin{figure}[t]
\centerline{\epsfysize=5.7in\epsffile{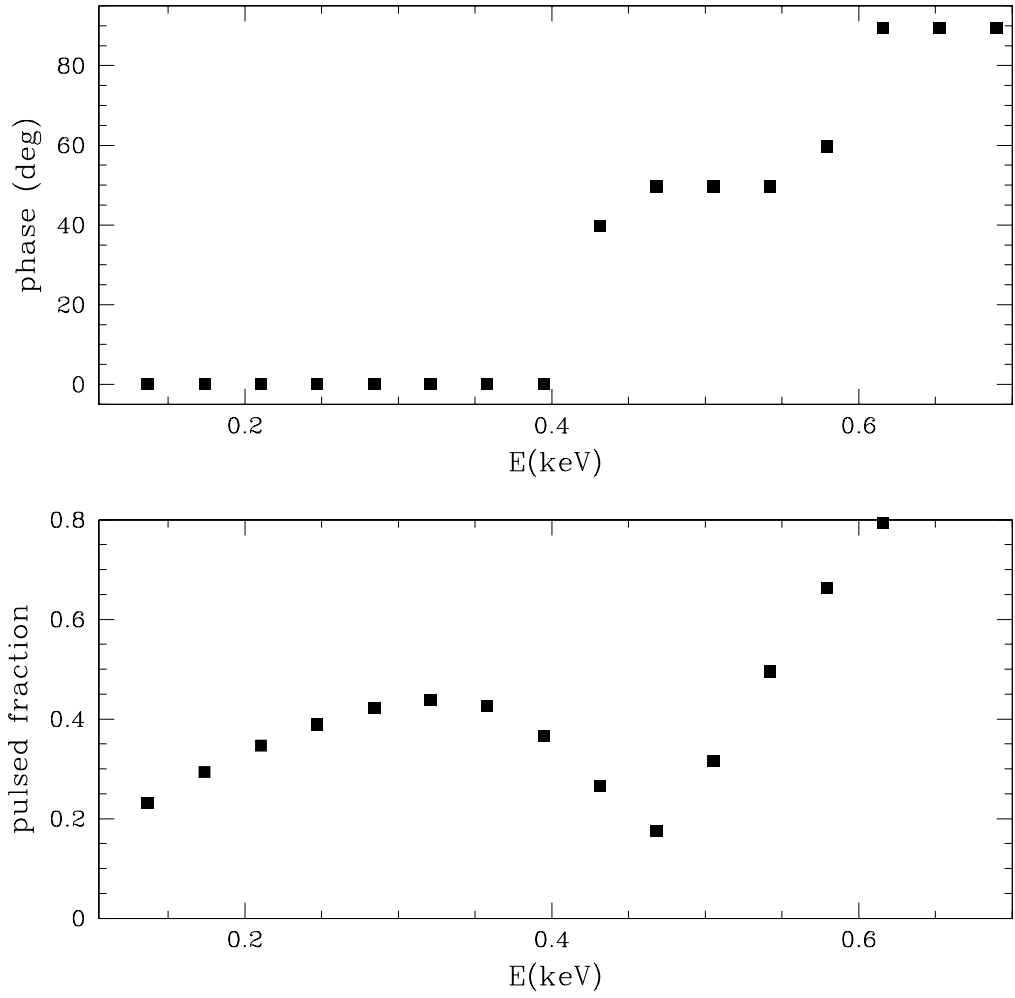}}
\caption{Predicted energy-dependent phase and pulsed fraction for
the X-ray emission from Geminga.}
\label{fig:3}
\end{figure}

\end{document}